\documentclass[aps,prl,twocolumn,groupedaddress,amsmath,amssymb,floatfix]
{revtex4}
\usepackage{graphicx,epsfig}

\begin{document}


\title{Scale invariant  extension of the standard model \\ with 
strongly interacting hidden sector}



\author{Taeil Hur}
\affiliation{School of Physics, KIAS, Seoul 130-722, Korea}


\author{P. Ko}
\email[]{pko@kias.re.kr}
\affiliation{School of Physics, KIAS, Seoul 130-722, Korea}


\date{\today}

\begin{abstract}
We present a scale invariant extension of the standard model 
with new QCD-like strong interaction in the hidden sector.  
A scale $\Lambda_H$ is dynamically generated in the hidden sector by 
dimensional transmutation, and chiral symmetry breaking occurs in the 
hidden sector. This scale  is transmitted to the SM 
sector by a real singlet scalar messenger $S$, 
and can trigger electroweak symmetry breaking (EWSB). 
Thus all the mass scales in this model arises from the hidden sector scale 
$\Lambda_H$ which has quantum mechanical origin. 
Furthermore the lightest hadrons in the hidden sector is stable by the 
flavor conservation of the hidden sector strong interaction, could be the 
cold dark matter (CDM). We study collider phenomenology, and relic 
density and direct detection rates of the CDM of this model. 
\end{abstract}


\maketitle



Understanding the origin of the electroweak symmetry breaking (EWSB) 
and the nature of cold dark matter (CDM) is the most important question  
in high energy physics in the Large Hadron Collider (LHC) era.
In particular, it is an open question whether all the masses of fundamental
particles can arise from new strong dynamics 
\cite{Hill:2005wg,Wilczek:2005ez,wilczek}.  
In our previous work \cite{koetal}, we considered a model with a  strongly 
interacting hidden sector, where a new strong interaction has such properties as 
confinement and chiral symmetry breaking like ordinary QCD: 
\begin{equation}
{\cal L}_{\rm hidden} = - {1\over 4} {\cal G}_{\mu\nu} {\cal G}^{\mu\nu}
+ \sum_{k=1}^{N_{h,f}} \left( \overline{\cal Q}_k i D\cdot \gamma
- M_{{\cal Q}_k} \right) {\cal Q}_k 
\end{equation}
plus interactions between the hidden sector and the SM 
sector due to some messengers which were not specified clearly.
Using the Gell-Mann-Levy's linear sigma model as a low energy effective 
theory for the hidden sector technicolor interaction, we made two 
important observations in Ref.~\cite{koetal}.  
The first point was that dynamical scale $\Lambda_H$, which is generated 
in the hidden sector by dimensional transmutation (alalogous to 
$\Lambda_{\rm QCD}$), can play a role of 
(or contribute to) the Higgs mass parameter in the SM.  
Another point was that the lightest mesons, the hidden sector pions 
can be good candidates for CDM. 
It is stable due to hidden sector flavor conservation, which is an accidental  
symmetry of hidden sector strong interaction. 
Note that we don't impose any ad hoc $Z_2$ symmetry for stability of CDM. 
However in Ref.~\cite{koetal}, the mass of the CDM (the hidden sector pion $\pi_h$) 
originated from explicit chiral symmetry breaking from nonzero current quark 
masses in the hidden sector [ $M_{{\cal Q}_k}$ in Eq. (1) ], and we did not 
ask the origin of those hidden sector current quark masses. 
It is important and interesting to investigate if the CDM mass can also 
arise from new strong dynamics through dimensional transmutation 
\cite{koetal}.

In this letter, we show that it is in fact possible to show that all the
mass scales including the CDM mass arise quantum mechanically through 
dimensional transmutation and the hidden sector chiral symmetry breaking. 
For that purpose, we start with classical lagrangian without any dimensionful 
parameters, which then possesses classical scale symmetry.  
The scale symmetry is broken only in the logarithm, which is related with 
the trace anomaly. 
For that purpose,  we have to introduce a real singlet 
scalar $S$ and modify the SM lagrangian as follows:
\begin{eqnarray}
  \label{eq:sm}
  {\cal L}_{\rm SM} & = & {\cal L}_{\rm kin} -
{\lambda_1 \over 2}~( H_1^{\dagger} H_1 )^2
- {\lambda_{1S} \over 2}~S^2 ~ H_1^{\dagger} H_1
- {\lambda_S \over 8}~S^4
\nonumber  \\
  & + &
\left( \overline{Q}^i H_1 Y^D_{ij} D^j 
+ \overline{Q}^i \tilde{H_1} Y_{ij}^U U^j
+ \overline{L}^i H_1 Y^E_{ij} E^j   \right.
\nonumber   \\
& + & \left. \overline{L}^i \tilde{H_1} Y^N_{ij} N^j
+ S N^{iT} C Y^M_{ij} N^j + h.c. \right)
\end{eqnarray}
The hidden sector lagrangian is also modified  as follows:
\begin{equation}
{\cal L}_{\rm hidden} = - {1\over 4} {\cal G}_{\mu\nu} {\cal G}^{\mu\nu}
+ \sum_{k=1}^{N_{h,f}} \left( \overline{\cal Q}_k i D\cdot \gamma
- \lambda_{k} S \right) {\cal Q}_k .
\end{equation}
Note that all the mass terms are replaced by $S$ or $S^2$ 
[including the mass  terms for the hidden sector quarks ($M_{{\cal Q}_i}$) 
by Yukawa coupling ($\lambda_{i} S$)] so that the classical lagrangian 
is scale invariant without any mass parameters at tree level. 
In our model, all the masses will be generated by quantum mechanical effects. 
Also $S$ will play a role of messenger connecting the SM and the hidden 
sectors. 

In fact the lagrangian (2)  was considered in  Ref.~\cite{Meissner:2006zh},  
where the radiative EWSB was studied a la Coleman and Weinberg \cite{coleman}.
In the following paper \cite{Meissner:2007xv}, the same authors showed that 
this lagrangian is renormalizable in a sense that the counterterm structure 
is the same as the classically scale invariant lagrangian when dimensional 
regularization and mass-independent renormalization scheme such as 
$\overline{MS}$ scheme are used.  
Our approach is different from Ref.~\cite{Meissner:2006zh} in that we include a 
hidden sector with new strong QCD-like interaction, so that we can discuss 
CDM and dynamical mass generation even without radiative corrections to the 
effective potential.

Note that the hidden sector has exact global 
$SU(N_{h,f})_L \times SU(N_{h,f})_R$ chiral symmetry in the limit $\lambda_k =0$
for $k=1,2,...N_{h,f}$, which will break into diagonal $SU(N_{h,f})_V$ with massless
Nambu-Goldstonte (NG) bosons. Due to strong interaction in the hidden sector, 
hidden sector quarks will condense with $\langle {\cal Q}_k {\cal Q}_k \rangle \neq 0$, 
and a linear term in $S$ will develops, leading to a nonzero VEV for $S$. 
After $S$ gets a VEV, one has $M_{{\cal Q}_i} = \lambda_i \langle S \rangle$.
The hidden sector quarks are light if 
$M_{{\cal Q}_j} \equiv \lambda_j \langle S \rangle \ll \Lambda_{h\chi} 
\equiv 4 \pi \Lambda_H$, where $\Lambda_{h\chi} $ is the hidden sector 
chiral symmetry breaking scale. 
If there are $N_f$ light hidden sector quarks, there would be approximate 
$SU( N_{h,f} )_L \times SU( N_{h,f} )_R$ global chiral symmetry.
This chiral symmetry is explicitly broken, so that the resulting hidden 
sector pions become massive with 
$m_{\pi_h}^2 = \mu_h M_{\cal Q} \ll \Lambda_{h\chi}$. 
If $n_{h,f}$ of $\lambda_i$'s are 
approximately equal, the $SU ( n_{h,f} )_V$ will be a good approximate 
global symmetry to categorize the pseudo NG bosons.

This picture is similar to SUSY models, where SUSY is broken spontaneously
by hidden sector gaugino condensation. The SUSY breaking effect is 
transimitted to the MSSM sector by messengers. Likewise,  in our model, 
classical scale symmetry is broken in the hidden sector by dimensional 
transmutation in our model, and its effect is transmitted to the SM sector 
by messenger (real singlet scalar $S$ in our model). 
Chiral symmetry breaking in the hidden sector is the same as 
the usual technicolor models, except that the hidden sector quarks are 
SM singlets rather than the SM doublets,  thereby the constraints from 
$S$ and $T$ become milder in our scenario.
We still do have a fundamental Higgs scalar $H_1$ in the SM sector, but
its mass parameter is determined by the hidden sector chiral symmetry 
breaking scale, and thus naturally suppressed relative to the Planck scale,
as the proton mass in QCD is naturally smaller than Planck scale by 
dimensional transmutation.

To illustrate our main points, let us consider $N_{h,f}=2$ with small 
current quark masses $M_{{\cal Q}_{i=1,2}}$ and $N_{h,c} = 3$, 
as in the ordinary QCD with two light flavors. 
In this case, the low energy degrees of freedom would
be the hidden sector pions ($\pi_h$) with decay constant 
$v_h \approx \Lambda_H$,  its scalar partner $\sigma_h$,
and the hidden sector nucleons $N_h = ( p_h , n_h )$.
To simplify our discussions, we assume that both hidden sector scalar 
$\sigma_h$ and nucleons $N_h$ are much heavier 
than the hidden sector pions, or the hidden sector scale $\Lambda_H$, 
and integrate them out. 
Then the low energy dynamics of the hidden sector 
can be described entirely in terms of the hidden sector pions.
This description is sufficient for  calculting the relic density and the
direct detection of $\pi_h$'s, since $\pi_h$'s are almost at rest in  
these calculations.  However, we will keep in mind that the hidden 
sector baryons would be also stable by hidden sector baryon number 
conservation, and make another CDM. Their relic density and direct 
detection rate is more difficult to calculate than those of hidden sector
pions, and so we do not include them in this letter. 

Within this picture, physics of the hidden sector pions and its interactions 
with the SM sector and the real scalar $S$ can be described by the 
following effective lagrangian:
\begin{eqnarray}
{\cal L}_{\rm hidden}^{\rm eff} &=& {v_h^2 \over 4} {\rm Tr} [
\partial_\mu \Sigma_h \partial^\mu \Sigma_h^\dagger ] 
+ {v_h^2 \over 2} {\rm Tr} [ {\bf \lambda} S
\mu_h ( \Sigma_h + \Sigma_h^\dagger ) ] 
\nonumber   \\
{\cal L}_{\rm mixing} & = & -v_h^2 \Lambda_{h\chi}^2 \left[ \kappa_H
{H_1^\dagger H_1 \over \Lambda_{h\chi}^2 } +\kappa_{S} {S^2 \over
\Lambda_{h\chi}^2} +\kappa'_{S} {S \over \Lambda_{h\chi}} \right.
\nonumber  \\
&+ & \left. O( {S H_1^\dagger H_1 \over \Lambda_{h\chi}^3} , { S^3
\over \Lambda_{h\chi}^3 }) \right]
\\
{\cal L}_{\rm full}&=& {\cal L}_{\rm hidden}^{\rm eff}+{\cal L}_{\rm SM}
+ {\cal L}_{\rm mixing}
\nonumber
\end{eqnarray}
In deriving ${\cal L}_{\rm mixing}$, we used the naive dimensional analysis,
and ignored openrators that are suppressed by powers of $v_h / \Lambda_{h\chi}$.
Dimensionless couplings $\kappa_H$, $\kappa_S$ etc. are of $O(1)$ according 
to the naive dimensional analysis.

Since we have to use an effective lagrangian for the hidden sector 
strong interaction, it is mandatory for us to discuss the effects of higher 
dimensional operators and power counting rules. Here one can adopt the 
naive power counting rule proposed by Georgi and Manohar 
\cite{Manohar:1983md}.
We have already showed some terms that connect the SM Higgs and $S$ 
to the hidden sector pions. We can also consider interactions involving 
the hidden sector pions and the SM fermions and gauge fields.
The basic principle is exactly the same as that in chiral lagrangian. 

The tree level scalar potential in the present model is given by  
\begin{eqnarray}
V&=& \frac{\lambda_1}{2} ( H_1^\dagger H_1 )^2 
+\frac{\lambda_{1S}}{2} H_1^\dagger H_1 S^2 + \frac{\lambda_S}{8} S^4 
\nonumber\\
&&+ \mu_{H_1}^2  H_1^\dagger H_1 + \frac{1}{2} \mu_S^2S^2 + \rho^3 S
\end{eqnarray}
where $\mu_{H_1}^2$, $\mu_S^2$ and $\rho^3$ are defined in terms of 
the parameters in the lagrangian:
\begin{eqnarray}
 \mu_{H_1}^2 &=& v_h^2 \kappa_H, \\
  \mu_{S}^2 &=& 2 v_h^2 \kappa_S,\\ 
  \rho^3 &=& v_h^2\{ \Lambda_{h\chi} \kappa'_S
  -\mu_h(\lambda_u + \lambda_d)\}.
\end{eqnarray}
We are interested in the phase: 
\[
H_1 = \left( 0 , {(v_1+h_{\rm SM})\over \sqrt{2}} \right)^T,~~~S = (v_S + S ).
\] 
The scalar mass matrix is given by  
(EWSB condition: $M_{11}+M_{22}>0, M_{11}M_{22}>M_{12}^2$)
\begin{equation}
{\cal L} \supset -\frac{1}{2} \begin{pmatrix}h_{\rm SM} & S \\
\end{pmatrix}
 \left(\begin{array}{cc} \lambda_1 v^2_1  & \lambda_{1S} v_1v_S \\
\lambda_{1S} v_1v_S & \lambda_S v^2_S -\rho^{3} /v_S \end{array}\right)
\begin{pmatrix} h_{\rm SM} \\ S \\
\end{pmatrix}
\end{equation}
The mass eigenstates $h$ and $H$ are defined as 
\begin{equation}
\left(\begin{array}{c} h \\ H\end{array}\right)=
\left(\begin{array}{cc}\cos\alpha & \sin\alpha \\ 
-\sin\alpha &\cos\alpha\end{array}\right)
\left(\begin{array}{c} h_{\rm SM} \\ S \end{array}\right) , 
\end{equation}
which diagonalizes the scalar mass matrix. 
The hidden sector pion get masses if $S$ develops a VEV: 
$m_{\pi_h}^2 = v_S \mu_h ( \lambda_u + \lambda_d)$. 


Our main point can be best illustrated for the simplest case, 
$\kappa_H= \kappa_S = \kappa_S' = 0$. In this case a reduction in the
number of parameters occurs, and we obtain 
$\lambda_{1S}$ and $\lambda_S$ in terms of $\lambda_1$ and $v_S$:
\begin{equation}
 \lambda_{1S} = - \frac{\lambda_1 v_1^2}{v_S^2}, ~~~
 \lambda_{S} = \frac{\lambda_1 v_1^4 + 2 m_{\pi_h}^2 v_h^2 }{v_S^4}.
 \end{equation}
Trading $\lambda_1$ with $m_h$, we use the following set as input 
parameters: $\tan \beta \equiv v_S/v_1$, $v_h$, $m_{\pi_h}$, $m_h$.  
For numerical analysis, we consider two cases: 
(a) $v_h = 500$ GeV and $\tan\beta = 1$ and (b) $v_h = 1000$ GeV and 
$\tan\beta = 5$, and scan over other two parameters with 
$m_{\pi_h} \lesssim 0.5 \Lambda_{h\chi} \sim 5 v_2$, 
so that the hidden sector pions can be still regarded as a pseudo 
Goldstone bosons. 
For $\lambda_1$, we scan upto $\sim 4 \pi$, and check 
if the perturbative unitarity for $W_L W_L$ scattering is satisfied.
The most important constraints on our model come from the Higgs
boson search and the relic density of the DM, the latter of which 
is calculated by modifying the micromega \cite{micromegas} suitable 
to our model.  


In the low energy, there are two scalars $h$ and $H$ which are linear 
combinations of $h_{\rm SM}$ and $S$, and both are assumed to be fundamental 
scalar bosons. 
Note that $h_{\rm SM}$ couples to the SM gauge bosons and SM fermions. 
Therefore the couplings of $h$ and $H$ are modified by $\cos\alpha$
and $\sin\alpha$ of the SM couplings. 
On the other hand the scalar $S$ couples to the SM Higgs boson,  the
right-handed Majorana neutrino through Majorana mass term, and most 
importantly to the hidden sector pions. In particular its coupling to the
hidden sector pion increases as the hidden sector pion mass increases,
which is characteristic of the model with classical scale symmetry. 
It is easy to copy the results from the SM case, with a few important 
things to be emphasized in our case.
$h (H) \rightarrow \pi_h \pi_h$ will open up the
invisible decay channel of Higgs bosons $h$ and $H$, whose
decay rates depend on the $\pi_h$ masses and other parameters.
This will make more difficult to observe the Higgs bosons $h$ or
$H$ using the visible final states. 
If $m_{H} > 2m_{h}$, then a new channel $H \rightarrow h h$ can open
up, and the decay width of $h_2$ will be increased. 
The production cross sections of $h$ and $H$ at the Tevatron or LHC 
are the same as the SM Higgs boson production rate, except that 
the rates are scaled by the $h(H)-t-t$ couplings, 
which are $\cos^2 \alpha$ and $\sin^2 \alpha$, respectively. 
Therefore the production rates of $h$ and $H$ are  always suppressed 
relative to the SM Higgs production of the same mass.

In Fig.~1, we show the branching ratio for $h$ with $m_h = 120$ GeV 
for (a) $v_h = 500$ GeV and $\tan\beta = 1$ and (b) $v_h = 1$ TeV and 
$\tan\beta = 2$. As in Ref.~\cite{koetal}, the invisible decay width of 
$h$ is large as long as $h \rightarrow \pi_h \pi_h$ is kinematically 
allowed ($m_{\pi_h} < m_h / 2 = 60$ GeV). Otherwise $h$ decay follows 
that of the SM predictions.

\begin{figure}
\includegraphics[width=7cm]
{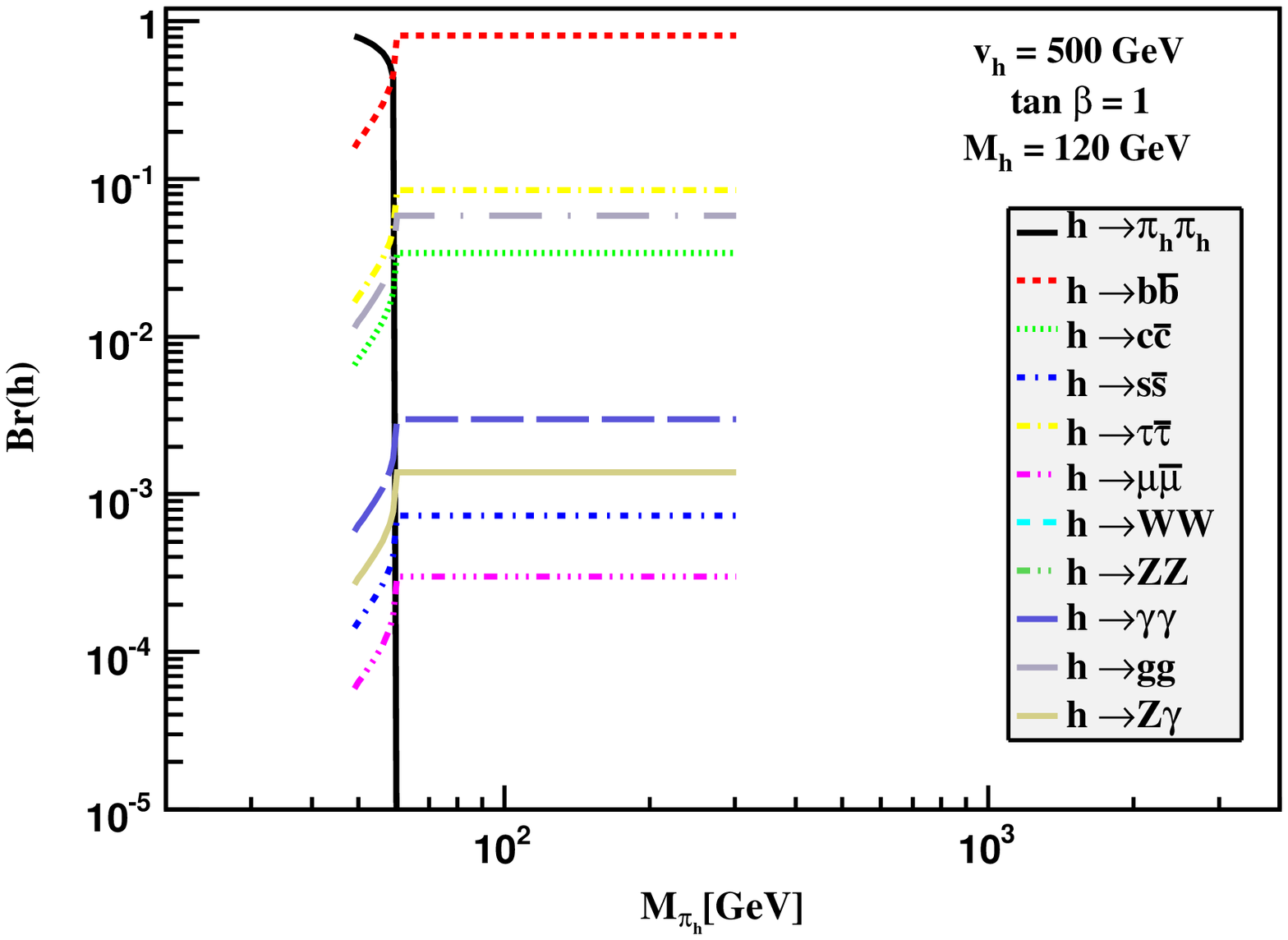} 
\includegraphics[width=7cm]
{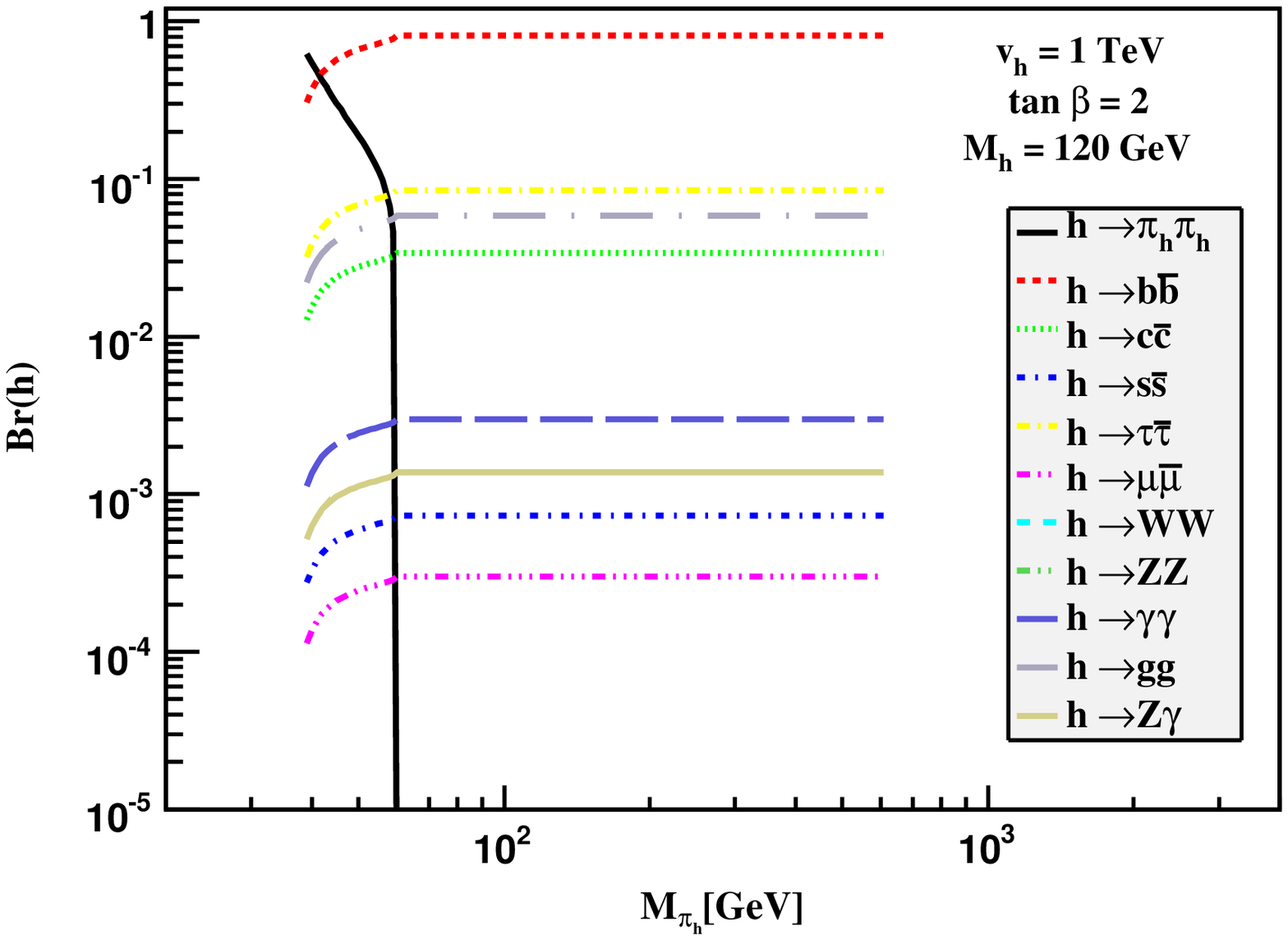} 
\caption{\label{fig:br-t1}
Branching ratios of $h$ of $m_h = 120$ GeV as 
functions of $m_{\pi_h}$ for (a)$v_h = 500$ GeV and $\tan\beta = 1$, 
and (b) $v_h = 1$ TeV and $\tan\beta =2$.} 
\end{figure}

The hidden sector pion $\pi_h$ can be a good candidate for the CDM
of the universe. 
In Fig.~2, we show the relic densities of $\pi_h$ in the 
$( m_{\pi_h} , m_h )$ plane for (a) $v_h = 500$ GeV and $\tan\beta = 1$,  
and (b) $v_h = 1$ TeV and $\tan\beta = 2$, 
with different colors for $\log_{10} (\Omega_{\pi_h} h^2)$ between 
$-4$ and $0.5$.  
We imposed only $\Omega_{\pi_h} h^2 < 0.122$  (the WMAP bound), since there 
could be additinoal DM's, namlely the lightest hidden sector nucleons or 
axions. The white region is excluded by direct search limit on Higgs boson 
and the correct EWSB vacuum for the SM sector. In a certain parameter 
space, the relic density is somewhat large, but not so large as in our
prevous paper \cite{koetal} 
where the hidden quark masses are given by hand, and not by scalar VEV. 

\begin{figure}
\includegraphics[width=7cm] 
{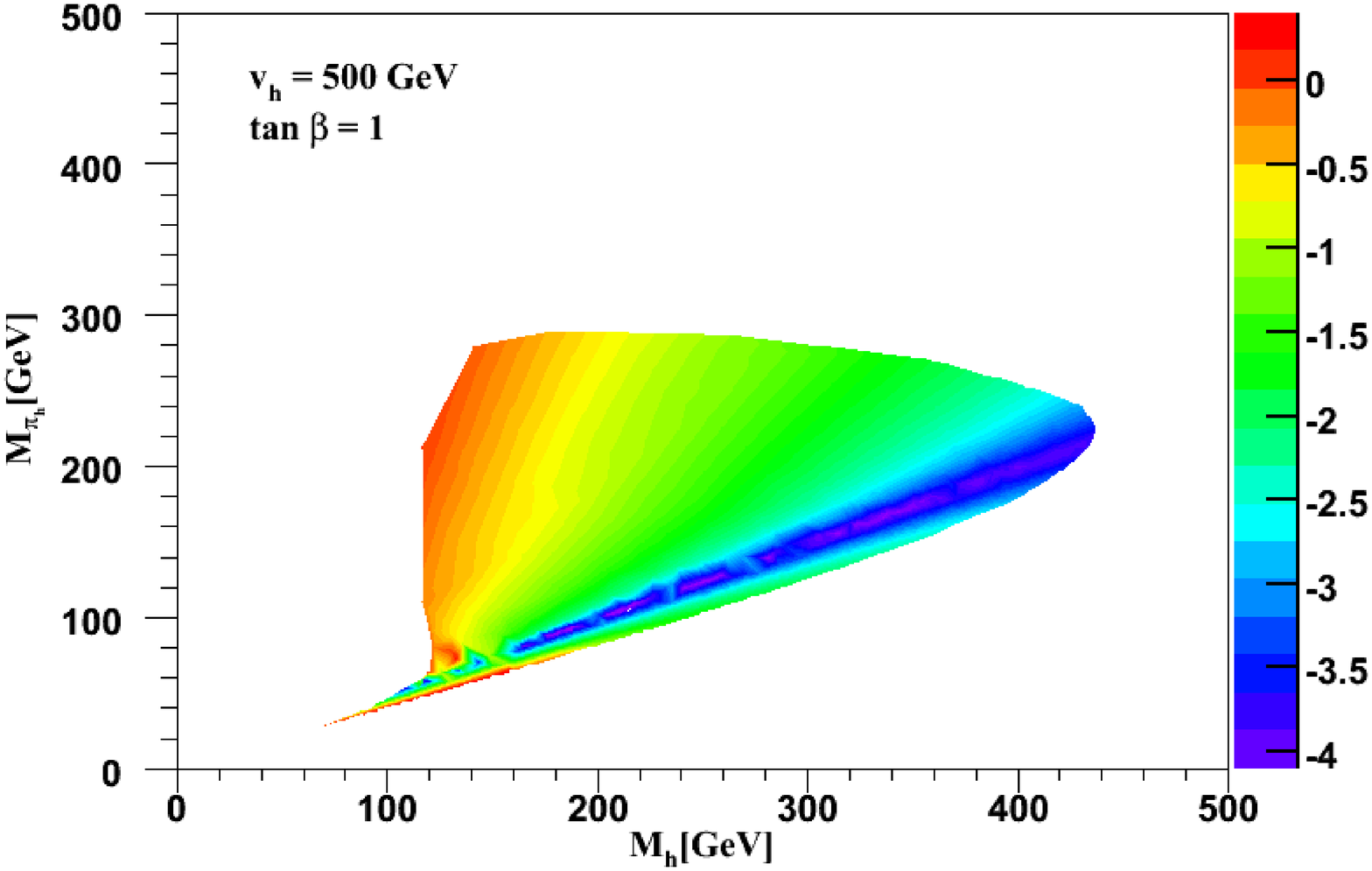}
\includegraphics[width=7cm] 
{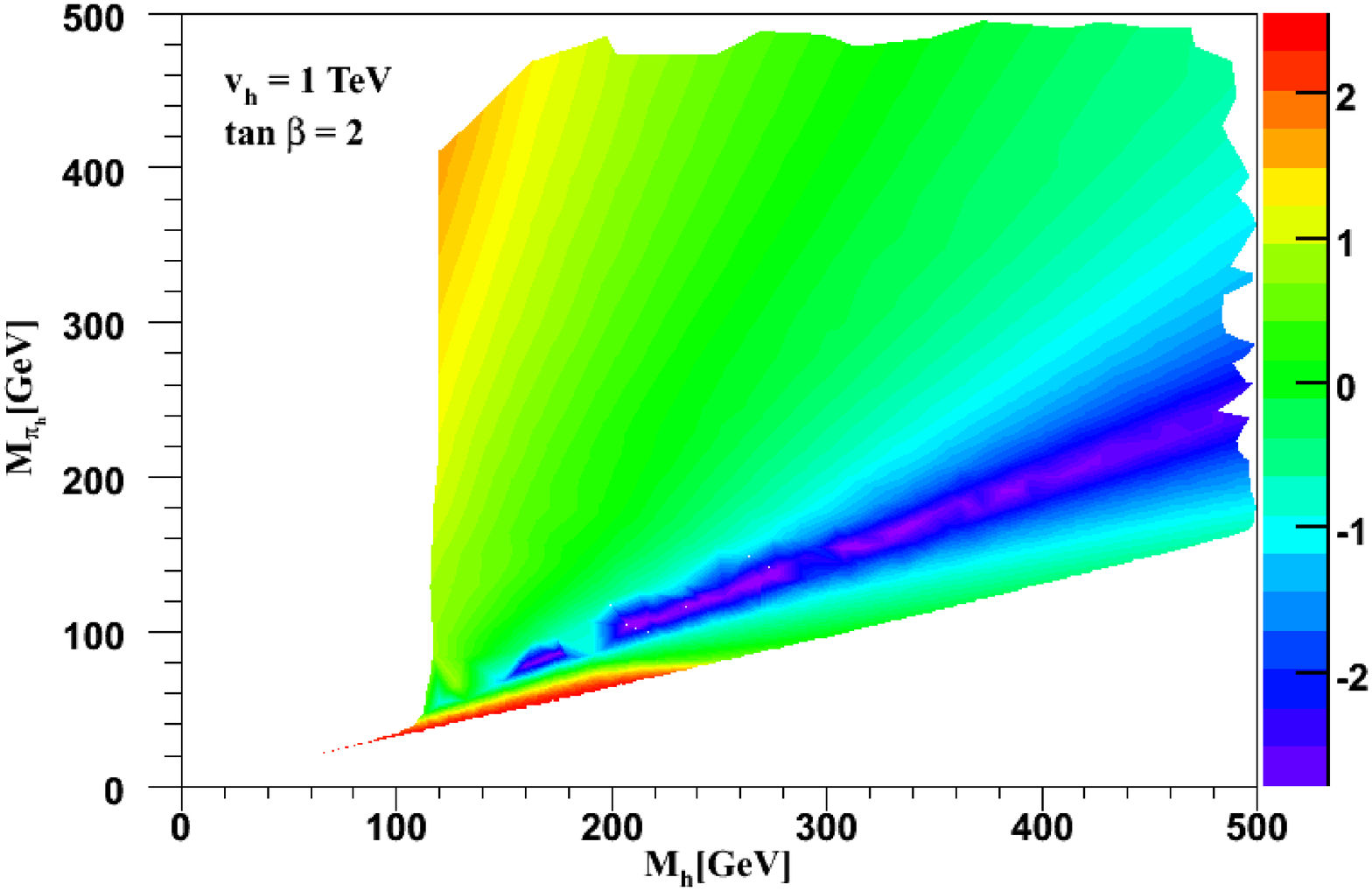}
\caption{
$\Omega_{\pi_h} h^2 $ in the $( m_{ h_1 } , m_{\pi_h} )$ plane  
for (a) $v_h = 500$ GeV and $\tan\beta = 1$, and 
(b) $v_h = 1$ TeV and $\tan\beta = 2$.} 
\end{figure}

In Fig.~3, we show the spin-independent dark matter scattering cross 
section with proton $\sigma_{\rm SI}$ for $( v_h , \tan\beta ) = 
(500 {\rm GeV}, 1)$, and $( 1 {\rm TeV}, 2)$, 
with the current bounds from CDMS-II and XENON 
as well as the projected sensitivities of XMASS and SuperCDMS. 
Within particular parameter space with all $\kappa$'s equal to zero, 
there is an approximate linear relationship between $\log \sigma_{\rm SI}$ 
and $\log M_{\pi_h}$. Blue and green dots denote the relic density 
$0.096 < \Omega_{\rm DM} h^2 < 0.122$ and $\Omega_{\rm DM} < 0.122$.
Note that our model predicts $\sigma_{\rm SI}$ just below the recent  
bound from XENON-10 experiment, and could be tested by the near future 
experiments. 

\begin{figure}
\includegraphics[width=7cm] 
{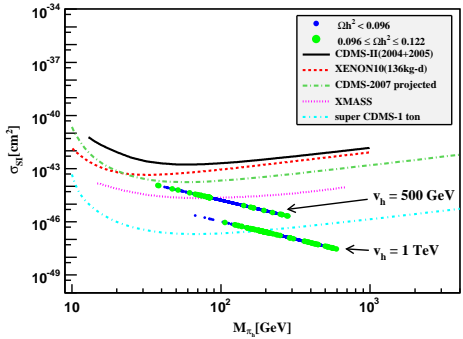}
\caption{\label{fig2}
$\sigma_{SI} (\pi_h p \rightarrow \pi_h p )$ as functions of $m_{\pi_h}$.  
The upper one is for $v_h = 500$ GeV and $\tan\beta = 1$, and 
the lower one is for $v_h = 1$ TeV and $\tan\beta = 2$.} 
\end{figure}


In this letter, we presented a renormalizable and scale invariant 
extension of the SM with new QCD-like strong interaction in the hidden sector. 
The scale generated in the hidden sector by dimensional transmutation 
is transmitted not only to the EWSB scale, but also to the hidden sector 
pion mass through $\lambda \langle S \rangle$. 
Therefore all the mass scales (except for the cosmological constant) 
have their origin in a single scale $\Lambda_H$ in the hidden sector 
strong interaction. 
A new element of the present work compared with our previous work 
\cite{koetal} was that the hidden sector quark masses also arise from 
scale invariant interaction and its breaking due to $\langle S \rangle \neq 0$. 
Further the present model is renormalizable, with $S$ being a messenger 
connecting the hidden sector and the SM sector.
As a concrete example to show this idea really works, we considered 
the $SU(2)_L \times SU(2)_R$ chiral symmetry breaking, and dynamics 
of the resulting two scalar bosons and the CDM ($\pi_h$'s) using 
effective chiral lagrangian technique. The details would depend on 
the number of hidden sector quark flavors, but the qualitative features
must be similar to our example based on two flavors in the hidden sector. 
Since we have assumed that $\langle S \rangle \lesssim 4 \pi v_h$, its scale is 
$\sim $ TeV. Therefore neutrinos get masses through TeV scale seesaw
mechanism, and the Majorana masses for $M_R$'s are $\langle S \rangle$
dependent. Interest in low energy seesaw models is being renewed these 
days for various reasons.

We can extend this work in various directions.
The lightest hidden sector baryons could be additional dark
matter candidate, and its phenomenology and relic density are interesting 
subjects.  We can consider an extra $U(1)_X$ gauge boson as a messenger, 
by introducing a new $U(1)_X$ under which both SM and hidden 
sector matters are charged \cite{Strassler:2006im,progress}.
Finally one can study the one loop radiative corrections a la Coleman 
and Weinberg \cite{coleman}. 
Such calculations have been done recently for the SM plus righthanded 
neutrinos by Meissner and Nicolai \cite{Meissner:2006zh}, 
and it was shown that the realistic EWSB could be possible. 
In their scenario, there is no CDM candidate, and it would be interesting
to extend their calcualtions to our model with vectorlike confining gauge 
theory in the hidden sector. 
More detailed discussions on the topics presented here will be
presented elsewhere \cite{progress}.

\begin{acknowledgments}
PK is grateful to W. Bardeen, M. Drees, C. Hill, D.W. Jung, J. Y. Lee, 
C.S. Lim, A. Masiero, S. Rudaz and F. Wilczek for discussions. 
PK is supported in part by National Research Foundation (NRF) 
through Korea Neutrino Research Center (KNRC) at Seoul National University.
\end{acknowledgments}

\end{document}